\documentclass[prl,aps,amsmath, graphicx, twocolumn]{revtex4}
\usepackage{graphicx}
\usepackage{amsmath}
\usepackage{amsfonts}
\begin{document}
%\draft
\title{Universal temperature dependence of the conductivity of a
strongly disordered granular metal}
\author{A.\,R.\,Akhmerov, A.\,S.\,Ioselevich\footnote{e-mail iossel@landau.ac.ru},}
\address{L. D. Landau Institute
for Theoretical Physics, 119334 Moscow, Kosygina 2, Russia}
\date{\today}

\begin{abstract}
A  disordered array of metal grains with large and random
intergrain conductances is studied within the one-loop accuracy
renormalization group approach. While at low level of disorder the
dependence of conductivity  on $\log T$ is nonuniversal (it
depends on details of the array's geometry), for
 strong disorder this dependence  is described by a
universal nonlinear function, which depends only on the array's
dimensionality. In two dimensions  this function is found
numerically. The dimensional crossover in granular films is
discussed.
\end{abstract}

\maketitle

{\bf Introduction}. Unusual temperature dependence of conductivity
$\sigma$ of granular metals has been in the focus of both
experimental (see, e.g., \cite{exp1,exp2,exp3} and references
therein) and theoretical
\cite{efetov,bel1,alt04,kam04,FIS,ZhangShklovskii,FI05,Kozub05,
bel-hopping2,Tran05} studies in recent years. In particular, the
logarithmic $T$-dependence of resistivity
\begin{equation}
\rho=A(1-\alpha\ln T), \label{logdep}
\end{equation}
was experimentally established in \cite{exp2} for low-resistivity
samples, while high resistivity samples, studied by the same
group, demonstrated the characteristic Efros-Shklovskii law
$\rho\propto \exp\{(T_{ES}/T)^{1/2}\}$ (see
\cite{FI05,bel-hopping2} for theoretical interpretation). The
empiric dependence \eqref{logdep} was found to be valid in the
wide $T$-range 30-300 K, where the resistivity $\rho$ varied by a
factor of order 3.

A similar weak $T$-dependence of low-resistivity granular films
was observed also in \cite{exp3}. Although in \cite{exp3} this
dependence was interpreted as a power-law $\rho=AT^{-\alpha}$ with
small $\alpha =0.117$, it could with the same accuracy be fitted
by the law \eqref{logdep}.

The  dependence \eqref{logdep} can not be explained by the usual
weak-localization effects because of many reasons. In particular,
the dependence \eqref{logdep} was apparently found in both two-
and three-dimensional samples, and, secondly, it appears to be
insensitive to the magnetic field.

The logarithmic $T$-dependence of {\it the conductivity} (not the
resistivity!) was theoretically derived by Efetov and Tschersich
\cite{efetov}. It arises due to  corrections to conductivity,
originating from local quantum fluctuations of intergrain
voltages, {\it $\grave{a}$ la} Ambegaokar, Eckern, and Sch\"{o}n
\cite{AES}. Under the conditions $g_0\gg 1$ (low resistivity at
high temperature) and $g_0\delta\ll T\ll E_C$ (moderately low
temperature), Efetov and Tschersich  have found that
\begin{equation}
\frac{\sigma(T)}{\sigma(T_R)}=1-\frac{4}{zg_0}\ln\frac{T_R}{T}
\label{ET}
\end{equation}
for an ideal lattice of identical grains with identical intergrain
conductances. Here $g_0$ is the dimensionless (measured in the
units of $e^2/2\pi\hbar$) intergrain conductance at high
temperature $T\sim T_R=g_0E_C$, $\delta$ is the intragrain level
spacing, $z$ is the coordination number of the lattice, and
$E_C\gg \delta$ is the charging energy of an individual grain.
Note, that $\tau_R\equiv \hbar/T_R$ is nothing else, but the
characteristic local charge relaxation time in the system. The
formula \eqref{ET} is valid for a lattice of arbitrary
dimensionality.

Later, in the paper \cite{FIS} the theory of Efetov and Tschersich
was generalized for the case of nonideal -- random -- array of
metal grains. It was shown, that the system of one-loop RG
equations, describing the renormalization of individual intergrain
conductances $g_{ij}$ upon lowering of temperature, can be written
in the form
\begin{equation}
\frac{d\ln g_{ij}}{dt}=-2R_{ij}(\{g_{kl}\}), \label{RG}
\end{equation}
where $t\equiv\ln(T_R/T)$ -- is the ``RG time'', and $R_{ij}$ are
dimensionless resistances of the entire network between the grains
$i$ and $j$. These resistances are some complicated nonlocal
functions of all the conductances in the array; to find them one
first would have to solve the  set of the Kirchhof equations
\begin{equation}
\sum_{l:\langle kl\rangle}
g_{kl}\left(V_k^{(ij)}-V_l^{(ij)}\right)=\delta_{ik}-\delta_{jk},
\label{kirchhof}
\end{equation}
 for voltages $V_k^{(ij)}$, and then find $R_{ij}(\{g_{kl}\})=V_i^{(ij)}-V_j^{(ij)}$.
This task is analytically not feasible for general random
inhomogeneous network of conductances.

For the case of weak  disorder it was shown in \cite{FIS} that the
theoretical dependence of $\sigma$ on $\ln T$ considerably
deviates from the linear law \eqref{ET} only in the vicinity of
the metal-insulator transition, where the role of disorder becomes
crucial. In the major part of the domain of the applicability of
the one-loop RG approach the Efetov-Tschersich law \eqref{ET}
remains a good approximation even for a  moderately strong
disorder, when the relative fluctuations of intergrain
conductances are of order of unity. At least this is true for the
case of square lattice, which was studied numerically in
\cite{FIS}.

In the present Letter we are considering the case of extremely
strong disorder. This study is motivated by two reasons. First,
the  Efetov-Tschersich linear law \eqref{ET} for the {\it
conductivity} is by no means identical to the experimentally
observed linear law \eqref{logdep} for {\it resistivity}. It is
important to understand, if the deviations from the law \eqref{ET}
due to the disorder can explain this discrepancy between theory
and experiment.

Secondly, one can expect  inhomogeneous fluctuations of
 $g_{ij}$ to be indeed very strong. The intergrain conductance
$g_{ij}$ exponentially depends on the thickness $d_{ij}$ of an
insulating layer between the grains
\begin{equation}
g_{ij} = \overline{g}_0e^{h_{ij}},\qquad h_{ij} =
-2\kappa(d_{ij}-\langle d\rangle), \label{distr}
\end{equation}
where $\overline{g}_0\equiv \exp\langle\ln g\rangle$. The
decrement of the electronic wave-function within the insulating
layer $\kappa\sim 1 - 5 \;\AA^{-1}$, so that already fluctuations
of $d_{ij}$ with quite moderate standard deviation
 of order of few Angstr\"{o}ms may cause gigantic fluctuations of
 $g_{ij}$. Thus, the condition of strong fluctuations
\begin{equation}
\Delta \equiv\sqrt{\langle h^2\rangle}\gg 1 \label{distr1}
\end{equation}
can be fulfilled very easily.

\begin{figure}
\includegraphics[width=\columnwidth]{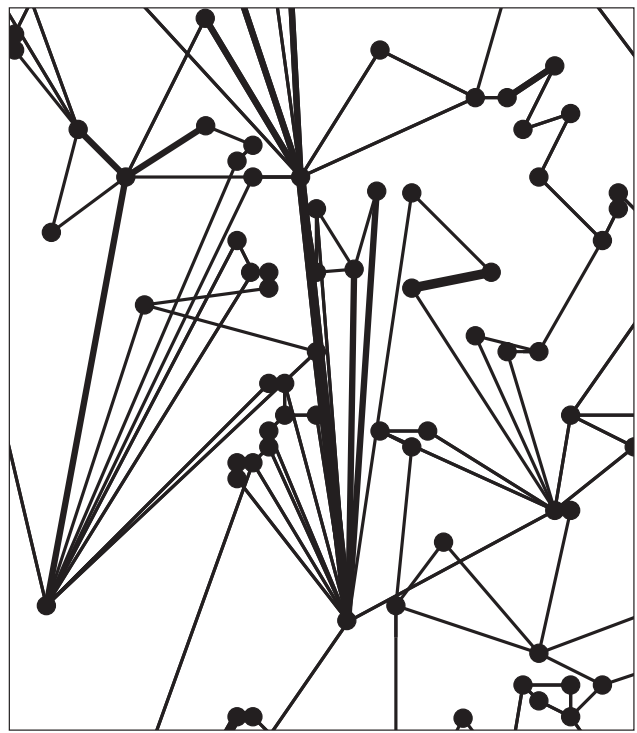}
\caption{A fragment of a critical network ${\cal
C}_{q=1.7}^{(2)}$, corresponding to a particular random
conductance network with $\Delta=8.66$ on a square lattice
$140\times 140$. The vertices (clusters) are shown as black
circles, a thickness of a bond between two vertices $n,n'$ is
proportional to the connection number $N_{nn'}$.} \label{backbone}
\end{figure}

{\bf Critical networks}. Under the condition \eqref{distr1}
neither the perturbation theory, nor the effective medium
approximation, used, respectively, in the case of weak and
moderately strong disorder in \cite{FIS}, can provide a reliable
tool for  finding the conductivity of the system. One can use
instead the theory of  highly inhomogeneous media (see, e.g.,
\cite{kirkpatrick,shklovskii}). According to this theory, the
conductivity of highly inhomogeneous network of conductances is
determined by the so called critical subnetwork, constructed
according to the following recipe (we will call it the
``$q$-procedure''):
\begin{itemize}
\item The critical conductance $g_c=\overline{g}e^{h_c}$ is found
from the condition
\begin{equation}
\int_{h_c}^{\infty}P(h)dh=p_c, \label{strong1}
\end{equation}
where $P(h)$ is the distribution function of $h$ (we assume that
there is no correlation between conductances of different
contacts), $p_c$ is the percolation threshold for the ``bonds
problem'' on a lattice of granules (regular, or irregular --
depending on the geometry of the array).

\item Certain number $q$ is chosen, such that
 $1\lesssim q\ll \Delta $.

 \item All the intergrain contacts with $h_{ij}<h_c-q$ are
substituted by infinite resistances (disconnected).

\item The contacts with $h_{ij}>h_c+q$ are substituted by ideal
conductances (abridged). Thus, the entire set of granules falls
apart into a number of {\it finite} clusters, all members of the
same cluster being ideally connected. These clusters serve as
vertices $n$ in the effective network.

\item The ``critical conductances'' with $|h-h_c|<q$ can either
connect granules within the same cluster $n$ (then they are
shunted by ideal conductors and can be discarded), or establish
links between different clusters $n$ and $n'$.  Thus, we arrive at
an effective network of vertices $n$, some of them being connected
by critical conductances. There may be more than one critical
conductance, connecting the same pair of vertices. By construction
(because $h_c+q>h_c$), besides finite  fragments, there must be
also an infinite connected subnetwork in this effective network,
spanning through the entire system. As usual, one can cut off all
``dangling ends'' of this infinite subnetwork and thus extract the
{\it backbone},  which is the substructure, relevant for the dc
conductivity (for the consistent definition of the backbone and
the dangling ends, see, e.g., \cite{bunde}). It is this backbone,
that we will call the $d$-dimensional
 {\it  critical network}  ${\cal C}_q^{(d)}$ of width $q$.
 A fragment  of a typical computer-simulated critical network is
shown in Fig.\ref{backbone}.
\end{itemize}

\begin{figure}
\includegraphics[width=\columnwidth]{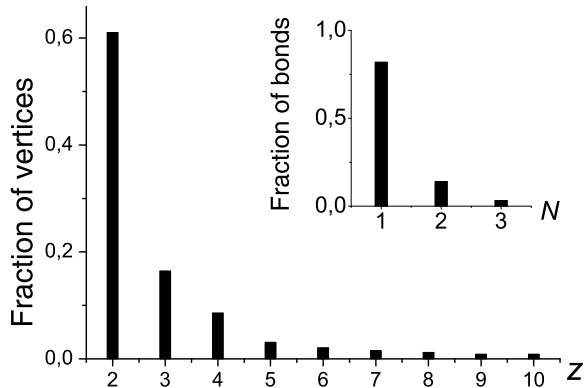}
\caption{The probability distributions for coordination numbers
$z$ and connection numbers $N$ for the same critical network, as
in Fig.\ref{backbone}.} \label{coordination-beg}
\end{figure}

Finally we are left with  a random multiply connected network, in
which every conductance is an independent random variable. The
geometry of this network is highly nontrivial. Distribution
functions for coordination numbers $z_n$, and for the ``connection
numbers'' $N_{nn'}$ (numbers of critical conductances, connecting
vertices $n$ and $n'$) are shown in Fig.\ref{coordination-beg}. It
should be noted that $z_n$ at different vertices, as well as
$N_{nn'}$ at different links, are correlated, because vertices,
corresponding to larger clusters of abridged grains,  have a
tendency to both larger coordination and larger connection
numbers.

The statistical geometry of the critical network, no matter how
complicated, is {\it universal}: it depends neither on details of
the initial lattice of grains, nor on the distribution function of
individual conductances. Topological properties of critical
networks for all systems of a given spatial dimension $d$ are
statistically equivalent. Different critical networks within the
same universality class differ from each other only in the spatial
scale -- the correlation radius (see \cite{shklovskii}):
\begin{equation}
\xi^{(d)}=a[P(h_c)]^{-\nu_d}\sim a\Delta^{\nu_d},\label{strong2aq}
\end{equation}
where $a$ is the typical intergrain distance, $\nu_d$ is the
critical exponent of the correlation radius for the percolation
problem in $d$ dimensions ($\nu_2=4/3$ and $\nu_3\approx 0.875$,
see \cite{bunde}). The conductivity of the critical network is
\begin{equation}
\sigma_q= \frac{e^2}{2\pi\hbar}A^{(d)}_q g_c
\left[\xi^{(d)}\right]^{2-d}\sim \frac{e^2a^{2-d}}{2\pi\hbar
a^{d-2}}\overline{g}_0\Delta ^{\nu_d(2-d)}e^{h_c},\label{strong2}
\end{equation}
where $A^{(d)}_q$ is a numerical prefactor.

Since the critical network ${\cal C}_q^{(d)}$ explicitly depends
on the value of the parameter $q$, so does the corresponding
conductivity $\sigma_q$. To get physically relevant
$\sigma_{q\to\infty}$ one has, in principle, to extrapolate the
result to $q\gg 1$. Fortunately, already for $q\gtrsim 1$ the
value of $q$ affects only the numerical preexponential factor
$A^{(d)}_q$, which rapidly converges to the true value $A^{(d)}$
as $q$ grows (the corrections are exponentially small at $q\gg
1$). The conductances with $|h_{ij}-h_c|\gg 1$ do not play any
considerable role in the critical subnetwork, therefore for a
qualitative discussion it is sufficient to choose some $q\sim 1$
and not to bother much about the effects of finite $q$. These weak
and purely quantitative effects will be discussed below, when we
will compare the numerical results, obtained for different values
of $q$.

{\bf Temperature dependence of the critical network's
conductivity}. In our problem the conductivity is also determined
by the critical network, but the conductances in this network
evolve with ``RG time'' $t$ according to the RG equations
\begin{equation}
\frac{d \ln g_{nn'}^{(k)}}{dt}=-2R_{nn'}, \label{RGa}
\end{equation}
where index $k=1,\ldots , N_{nn'}$ distinguishes different
conductances, connecting the same pair $nn'$ of vertices;
$R_{nn'}$ is the resistance of the critical network between
vertices $n$ and $n'$. It is very important, that these
resistances are governed solely by the critical conductances, so
that the system \eqref{RGa} is closed.

Let us introduce normalized variables
\begin{equation}
y_{nn'}^{(k)}=h_{nn'}^{(k)}-h_c,\quad
\tilde{t}=\frac{t}{A^{(d)}g_c},\quad
\tilde{\sigma}=\frac{\sigma}{A^{(d)}g_c}\left[\xi^{(d)}\right]^{d-2}\frac{2\pi\hbar}{e^2},\label{strong3}
\end{equation}
then
\begin{itemize}
\item The critical network is reduced to a completely universal
one, with the spatial scale equal to unity. \item The equations
\eqref{RG}  take the universal form
\begin{equation}
\frac{dy_{nn'}^{(k)}}{d\tilde{t}}=-2A^{(d)}\tilde{R}_{nn'}\left(\left\{
e^{y_{ll'}^{(p)}}
 \right\}\right). \label{RGtilde}
\end{equation}
\item Initial
 conditions $y_{nn'}^{(k)}(\tilde{t}=0)$ for the RG equations \eqref{RGtilde} independently take random
 values with a universal distribution function
\begin{equation}
\tilde{P}_q(y)=\frac{\theta(q-|y|)}{2q}. \label{distrib1}
\end{equation}
\end{itemize}
Thus we have arrived at a completely universal mathematical
problem: One has to solve a set of universal equations (with
universal initial conditions) for variables $y_{nn'}^{(k)}$,
defined on a random graph with universal statistical properties.
 As a result, the normalized conductivity
$\tilde{\sigma}$ should be a {\it universal function} of the
normalized time $\tilde{t}$:
\begin{equation}
\tilde{\sigma}_q=F^{(d)}_q(\tilde{t}),\label{strong4}
\end{equation}
the shape of this function depends only on the space
dimensionality $d$ and on the parameter $q$. At $q\gtrsim 1$ the
$q$ dependence is very weak and can be neglected, therefore  we
omit the index $q$ in what follows. Unfortunately, we were not
able to find this universal function analytically, so we had to do
that numerically (see below).

Coming back to the initial physical variables, we get the global
conductivity of the system in the form:
\begin{equation}
\frac{\sigma(T)}{\sigma(T_R)}=
F^{(d)}\left[\frac{e^2/2\pi\hbar}{\left[\xi^{(d)}\right]^{d-2}\sigma(T_R)}\ln\frac{T_R}{T}\right],\label{strong5a}
\end{equation}
where  $T_R\equiv g_cE_C$.  The only nonuniversal (i.e., depending
on the details of the conductances' distribution function)
ingredient in this formula is $\xi^{(d)}\sim \Delta^{\nu_d}$. Note
also, that there is absolutely no reminiscence of the nonuniversal
short-scale geometric structure of the array.

{\bf Lateral conductance of granular films}. One should
distinguish between thin films, whose thickness $\delta L\ll
\xi^{(3)}$, and thick films with $\delta L\gg \xi^{(3)}$. The
arguments of paper \cite{shkl1} (see also book \cite{shklovskii}),
designed for the description of hopping conductivity in films, can
be modified for our case. The true 3-dimensional behavior takes
place only in thick films, where the dimensionless conductance
 of a square sheet
of the film is $G_{\square}=\sigma^{(3)}\delta L$, and
\begin{equation}
\frac{G_{\square}(T)}{G_{\square}(T_R)}=
F^{(3)}\left[\left(\frac{\delta L}{\xi^{(3)}}\right)\frac{\ln
(T_R/T)}{G_{\square}(T_R)}\right].\label{strong5a}
\end{equation}

For thin films
\begin{equation}
\frac{G_{\square}(T)}{G_{\square}(T_R)}= F^{(2)}\left[\frac{\ln
(T_R/T)}{G_{\square}(T_R)}\right],\label{strong5p}
\end{equation}
and $G_{\square}$ exponentially depends on $\delta L$:
\begin{equation}
G_{\square}(T_R)=A^{(d)} \overline{g}_0e^{h_c(\delta L)},
\label{strong5bi}
\end{equation}
where $h_c=h_c(\delta L)$ is determined as a threshold of  lateral
percolation in a slab of thickness $\delta L$ cut out of a
three-dimensional network of all conductances with $h_{ij}<h_c$.
The threshold condition has a form
\begin{equation}
\xi^{(3)}(h_c)= c\;\delta L, \label{stro}
\end{equation}
where the three-dimensional correlation length $\xi^{(3)}(h_c)\sim
a [(h_c^{(3)}-h_c)/\Delta]^{-\nu_3}$, and $c\sim 1$. As a result,
in the
 range  $a\ll \delta L\ll \xi^{(3)}$ one finds
\begin{equation}
h_c^{(3)}-h_c(\delta L)=D\Delta (a/\delta L)^{1/\nu_3},
\label{strong5bi}
\end{equation}
where $D$ is some nonuniversal numerical constant (see
\cite{shklovskii}). Note, that this formula
 interpolates between $h_c^{(3)}$ in the case of a
bulk, thick film, and $h_c^{(2)}$ in the case of a one-monolayer
film.

The  quasi two-dimensional critical network $\tilde{\cal
C}_q^{(2)}$, arising in the film as a result of the $q$-procedure,
centered at $h_c(\delta L)$, is characterized by the effective
two-dimensional correlation length $\xi_{\rm eff}^{(2)}$. To
estimate this length we write it in a form $\xi_{\rm
eff}^{(2)}=a(\delta L/a)^{\alpha_1}\Delta^{\alpha_2}$, which
follows from the scaling arguments. This expression, valid in the
range $a\ll\delta L\ll \xi^{(3)}$, should match with the true
two-dimensional behavior $\xi_{\rm eff}^{(2)}=\xi^{(2)}\sim
\Delta^{\nu_2}$ at $\delta L\sim a$, and with the true
three-dimensional one $\xi_{\rm eff}^{(2)}=\xi^{(3)}\sim
\Delta^{\nu_3}$ at $\delta L\sim \xi^{(3)}$. These two matching
conditions fix the exponents $\alpha_1,\alpha_2$, so that we find
$\alpha_1=1-(\nu_2/\nu_3)$ and $\alpha_2=\nu_2$. As a result,
$\xi_{\rm eff}^{(2)}$ can be conveniently written in a form
\begin{equation}
\xi_{\rm eff}^{(2)}\sim \delta L \left(\xi^{(3)}/\delta
L\right)^{\nu_2/\nu_3}\gg \delta L. \label{strong500}
\end{equation}

 Thus, our quasi-two-dimensional critical network is indeed effectively {\it strictly
two-dimensional}, hence it belongs to the corresponding $d=2$
universality class, and the universal function is $F^{(2)}$ for
all films with $\delta L\ll \xi^{(3)}$ (cf. formula
\eqref{strong5p}), not only for those, consisting of just one
monolayer of grains.

{\bf Numerical results}.  In this Letter we present the numerical
results only for the two-dimensional case. For determination of
the function $F^{(2)}_q(\tilde{t})$ we have generated random
$L\times L$ square arrays of conductances with distribution
$P(h)=\theta(\sqrt{3}\Delta-|h|)/2\sqrt{3}\Delta$ for different
$20<L<140$ and $0<\Delta<8.66$. For the samples with large
$\Delta$ we constructed  critical networks ${\cal C}_q^{(2)}$ with
$0.5<q<2$,
 and numerically solved the system of
equations \eqref{RGtilde} and \eqref{kirchhof} on these networks,
which are very much smaller arrays, than the initial ones. Note,
that only due to this truncation of the array, as well as to the
fact, that the conductances of the critical network are all of the
same order of magnitude, the problem becomes feasible. The
numerical solution of the full set of Kirchhof equations for a
reasonably large array of strongly different conductances would be
practically impossible. Thus, the concept of the critical network
turns out to be crucially helpful not only for understanding of
basic physics, but also for the numerics.

\begin{figure}
\includegraphics[width=1.0\columnwidth]{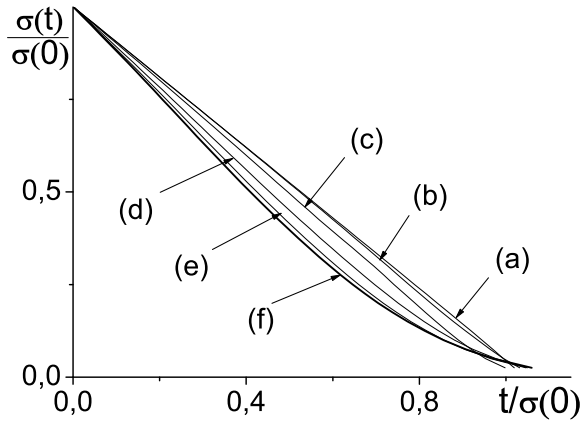}
\caption{Dependence of the normalized conductance per square for a
thin film on the parameter $t=\ln(T_R/T)$ for different standard
deviations $\Delta$ of the logarithms of intergrain conductances.
$M$ is the number of samples in the averaging ensemble. In all
cases $q=1.7$ and $L\gg \xi^{(2)}(\Delta)$. (a) -- $\Delta=0.14$,
$L=50$, $M=150$; (b) -- $\Delta=0.58$, $L=50$, $M=150$; (c) --
$\Delta=1.16$, $L=50$, $M=300$; (d)  -- $\Delta=2.31$, $L=80$,
$M=300$; (e)  -- $\Delta=3.47$, $L=100$, $M=400$; (f)  --
$\Delta=8.66$, $L=140$, $M=400$. in all cases $q=1.7$ and $L\gg
\xi^{(2)}(\Delta)$. The thick line -- universal function
$F^{(2)}(\tilde{t})$.} \label{range}
\end{figure}

The main result -- the dependence of the conductivity on $\ln
(T_R/T)$ for different values of $\Delta$ is shown in
Fig.\ref{range}. One can see that this dependence becomes
universal already at $\Delta\sim 3 - 4$: the difference between
curves with $\Delta=3.47$ and  $\Delta=8.66$ is negligible, so we
can with reasonably high accuracy identify the last one with the
universal function $F^{(2)}(\tilde{t})$. In the range
$0<\tilde{t}<0.5$ it is approximately linear with a slope,
corresponding to an effective coordination number $z_{\rm
eff}\approx 3.2$, which is consistent with the statistics of
coordination, shown in Fig.\ref{coordination-beg}. Note, that
$z_{\rm eff}$ is universal and does not depend on the coordination
number $z$ of the initial lattice. At $\tilde{t}>0.5$ the slope
starts to decrease rapidly, and $F^{(2)}(\tilde{t})$ goes to zero
at $\tilde{t}_c\approx 1.2$. Unfortunately, the problem of finding
the critical exponent, which characterizes the behavior of
$F^{(2)}(\tilde{t})$ at $\tilde{t}\to\tilde{t}_c$, seems to be
numerically not feasible.

\begin{figure}
\includegraphics[width=0.9\columnwidth]{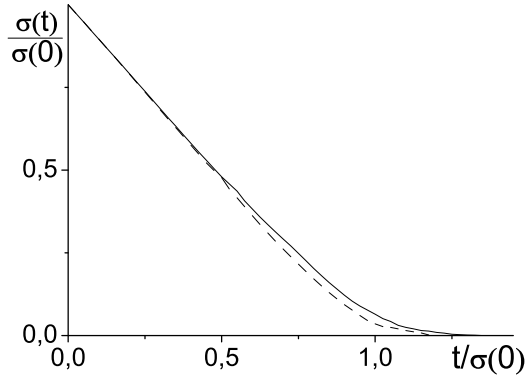}
\caption{Comparison of the   solutions of the full set of equation
for the $40\times 40$ sample with $\Delta= 1.734$ (solid line) and
the set, truncated according to the $q$-procedure with $q=1.7$
(broken line).} \label{comparison}
\end{figure}

We also have checked the accuracy of the method from the point of
view of finite size and finite $q$ corrections. For an array
$40\times 40$ with $\Delta=1.734$ we were able to perform the
straightforward calculations (involving the entire array) and
compare their results to those, obtained on truncated ${\cal
C}_q$-networks. The results of this comparison, shown in
Fig.\ref{comparison}, demonstrate a very good accuracy already at
moderate values of $q=1.7$. This accuracy is reached at the
network, which includes only about a half of all conductances of
the array. Note, that although the  condition $q\ll\Delta$,
required for general distributions $P(h)$, is obviously not
satisfied in this particular simulation, for the rectangular shape
of $P(h)$, used here, the strong inequality can be substituted by
a milder condition $q<\sqrt{3}\Delta$.

\begin{figure}
\includegraphics[width=0.9\columnwidth]{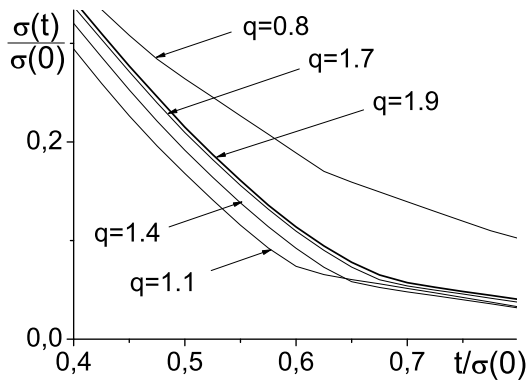}
\caption{Comparison of the solutions, using $q$-procedure with
different $q$. One can see that the difference between $q=1.7$ and
$q=1.9$ results is already negligible.} \label{q-dependence}
\end{figure}

For a sample $140\times 140$ with $\Delta=8.66$ (the same
realization, as used in
Figs.\ref{backbone},\ref{coordination-beg}) we performed the
calculations with different values of $q$. The results, shown in
Fig.\ref{q-dependence} demonstrate fast convergence of the
procedure: the difference between $F_q^{(2)}$ for $q=1.7$ and
$q=1.9$ is already very small, so that using $q=1.7$ procedure for
evaluating the universal function $F^{(2)}$ is a very good
approximation.

\begin{figure}
\includegraphics[width=0.9\columnwidth]{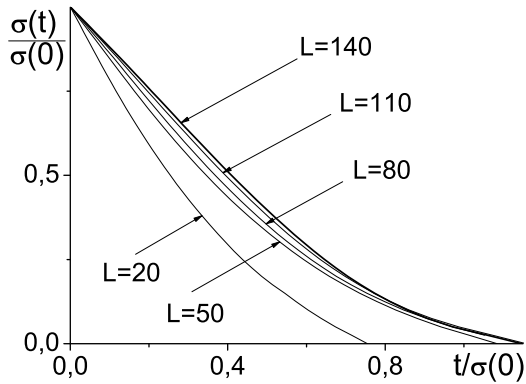}
\caption{Comparison of solutions  for different lattice sizes $L$.
All curves are obtained for largest standard deviation
$\Delta=8.66$ by the $q$-procedure with $q=1.7$, and averaged over
$M=400$ realizations.} \label{size}
\end{figure}

From general arguments it is clear that to get self-averaging
results, one should use  arrays with size $L> \xi^{(2)}$. In
Fig.\ref{size} we have plotted functions $F^{(2)}(\tilde{t})$,
found for arrays with  the same $\Delta =8.66$, but with different
sizes $L$. For these arrays one expects $\xi^{(2)}\sim
C[P(h_c)]^{-\nu_2}=C\cdot 30^{4/3}=93C$ (with unknown numerical
coefficient $C$). We see that for $L>100$ the curves become almost
identical, so  we conclude that $C\sim 0.5 - 1$ and square arrays
with $L>[P(h_c)]^{-\nu_2}$ can empirically be considered as large
enough.


 In principle, the proper thing to do would be an accurate extrapolation of  finite-$q$ and finite-$L$ results
  to $q=\infty$ and $L=\infty$.
 Unfortunately, the complexity of the numerical solution at large
 $\Delta$, $q$, and $L$ did not allow us to reach the accuracy, necessary to
 such an extrapolation.
The larger $q$ is chosen for $q$-procedure, the less stable and
precise the calculation of $R_{ij}$ becomes. This instability
limits $q$ by about $q=2$. Another precision limitation  is  rapid
increase of computational complexity with increase of
$L/\xi^{(2)}$. The mentioned difficulties limit the accuracy of
each curve by about 5\%, which is comparable with the separation
of neighboring curves on our plots, making the extrapolation
procedure senseless. However, since  a rapid convergence of the
results is observed, we believe that the universal curve
$F^{(2)}(\tilde{t})$ lies within the same 5\% error interval near
the curve with $\Delta=8.66$, $L=140$.

{\bf Conclusions}. Thus, the dependence of conductivity $\sigma$
on $\ln T$ for strongly disordered arrays of grains is nonlinear
and {\it universal}, contrary to the case of a regular array, for
which this dependence is linear, with a nonuniversal slope. The
sign of the curvature of this nonlinear function is such, that the
effects of strong disorder seem to improve the agreement between
the theory and the experimental law \eqref{logdep}. Unfortunately,
the available experimental data
 is still not sufficient for a direct quantitative  comparison
 of theory and experiment. In particular, it is hard to make reliable estimates for $T_R$
 and $G_{\square}(T_R)$. It is also worth reminding that the
 one-loop RG, which is the basis of the present theory, is valid
 at temperatures well above the metal-insulator transition.

 The universality of the temperature
 dependence of the conductivity is the direct consequence of the
 universality of the critical network, which carries the current
 in a highly inhomogeneous system. This critical
 network in a granular film appears to be effectively two-dimensional, if the
 film's thickness $\delta L$ is less than the three-dimensional
 correlation length $\xi^{(3)}\sim a\Delta^{\nu_3}$, the latter
 being large at high level of
 disorder.

We are grateful to M.V.Feigelman, D.S.Lyubshin and M.A.Skvortsov
for useful discussions. Special thanks are due to D.S.Lyubshin for
his invaluable advises, concerning the numerical aspect of this
work. The project was supported by the Program ``Quantum
Macrophysics'' of RAS and by RFBR grant 06-02-16533.

\end{document}